\documentclass[10pt, conference]{IEEEtran}
\IEEEoverridecommandlockouts
\usepackage{cite}
\usepackage{amsmath,amssymb,amsfonts}
\usepackage{algorithm}
\usepackage{algpseudocode}
\usepackage{graphicx}
\usepackage{textcomp}
\usepackage{xcolor}
\usepackage{enumitem}
\usepackage{tikz}
\usepackage[acronym,toc]{glossaries}
\usepackage{color,soul}
\usepackage[caption=false,font=normalsize,labelfont=sf,textfont=sf]{subfig}
\usepackage[acronym,nonumberlist]{glossaries-extra}
\usepackage{multicol,multirow}
\usepackage{booktabs}
\usepackage[table]{xcolor}
\usepackage{layout}
\makeatletter
\IEEEsettextheight{0.81in}{1.03in}
\IEEEsettopmargin{t}{0.81in}
\IEEEquantizetextheight{c}
\IEEEsettopmargin{q}{0sp}
\makeatother
\makeglossaries
\setabbreviationstyle[acronym]{long-short}

\newacronym{2d}{2D}{two-dimensional}
\newacronym{3d}{3D}{three-dimensional}
\newacronym{3gpp}{3GPP}{3rd generation partnership project}
\newacronym{3msk}{3MSK}{3-level minimum-shift-keying}
\newacronym{4g}{4G}{fourth-generation}
\newacronym{5g}{5G}{fifth-generation}
\newacronym{6dof}{6DoF}{six degrees of freedom}
\newacronym{6g}{6G}{sixth-generation}
\newacronym{6tisch}{6TiSCH}{IPv6 over the TSCH mode of IEEE 802.15.4e}
\newacronym{abp}{ABP}{authentication by personalisation}
\newacronym{ac}{AC}{alternating current}
\newacronym{aci}{ACI}{adjacent channel interference}
\newacronym{aclr}{ACLR}{adjacent channel leakage ratio}
\newacronym{acpr}{ACPR}{adjacent channel power ratio}
\newacronym{adc}{ADC}{analog-to-digital converter}
\newacronym{adr}{ADR}{adaptive data rate}
\newacronym{aec}{AEC}{aluminum electrolytic capacitor}
\newacronym{aes}{AES}{Advanced Encryption Standard}
\newacronym{afe}{AFE}{analog front end}
\newacronym{ag}{AG}{array gain}
\newacronym{agc}{AGC}{automatic gain controller}
\newacronym{agps}{A-GPS}{Assisted Global Positioning System} 
\newacronym{ai}{AI}{artificial intelligence}
\newacronym{aimd}{AIMD}{Additive-Increase/Multiplicative-Decrease}
\newacronym{aiot}{A-IoT}{Ambient IoT}
\newacronym{alk}{ALK}{Alkaline}
\newacronym{am}{AM}{amplitude modulation}
\newacronym{amam}{AM/AM}{amplitude modulation to amplitude modulation}
\newacronym{ambc}{AmBC}{ambient backscatter communication}
\newacronym{amc}{AMC}{automatic modulation classification}
\newacronym{amp}{AMP}{AMbient Power}
\newacronym{ampm}{AM/PM}{amplitude modulation to phase modulation}
\newacronym{aoa}{AOA}{angle-of-arrival}
\newacronym{aod}{AOD}{angle-of-departure}
\newacronym{aoi}{AoI}{age of information}
\newacronym{ap}{AP}{access point}
\newacronym{api}{API}{application program interface}
\newacronym{apt}{APT}{acoustic power transfer}
\newacronym{apu}{APU}{access point unit}
\newacronym{ar}{AR}{augmented reality}
\newacronym{arima}{ARIMA}{Auto-Regressive Integrated Moving Average}
\newacronym{arp}{ARP}{Antenna Reference Point}
\newacronym{asic}{ASIC}{application specific integrated circuit}
\newacronym{ask}{ASK}{amplitude-shift keying}
\newacronym{at}{AT}{ATtention}
\newacronym{auv}{AUV}{autonomous underwater vehicle}
\newacronym{awg}{AWG}{arbitrary waveform generator}
\newacronym{awgn}{AWGN}{additive white Gaussian noise}
\newacronym{baw}{BAW}{bulk acoustic wave}
\newacronym{bb}{BB}{base-band}
\newacronym{bc}{BC}{backscatter communication}
\newacronym{bcjr}{BCJR}{Bahl-Cocke-Jelinek-Raviv}
\newacronym{bd}{BD}{backscatter device}
\newacronym{be}{BE}{Belgium}
\newacronym{ber}{BER}{bit error rate}
\newacronym{bf}{BF}{beamforming}
\newacronym{bga}{BGA}{ball grid array}
\newacronym{bjt}{BJT}{bipolar junction transistor}
\newacronym{bldc}{BLDC}{brushless DC}
\newacronym{ble}{BLE}{Bluetooth low energy}
\newacronym{bler}{BLER}{block error rate}
\newacronym{bod}{BOD}{Brown-Out Detection}
\newacronym{bom}{BOM}{bill of materials}
\newacronym{bpsk}{BPSK}{binary phase-shift keying}
\newacronym{brp}{BRP}{beam refinement process}
\newacronym{bs}{BS}{base station}
\newacronym{btwt}{bTWT}{broadcast TWT}
\newacronym{bu}{BU}{booster unit}
\newacronym{bw}{BW}{bandwidth}
\newacronym{ca}{CA}{carrier emitter}
\newacronym{cad}{CAD}{channel activity detection}
\newacronym{cars}{CARS}{calibration reference signal}
\newacronym{cbm}{CBM}{condition based maintenance}
\newacronym{cbra}{CBRA}{contention-based random access}
\newacronym{cc}{CC}{constant current}
\newacronym{ccdf}{CCDF}{complementary cumulative distribution function}
\newacronym{ccnn}{CCNN}{circular convolutional neural network}
\newacronym{ccs}{CCS}{correlative channel sounder}
\newacronym{cdf}{CDF}{cumulative distribution function}
\newacronym{cdma}{CDMA}{code division-multiple access}
\newacronym{cdrx}{CDRX}{connected mode DRX}
\newacronym{ce}{CE}{coverage enhancement}
\newacronym{ced}{CED}{cumulative energy density}
\newacronym{ceofdm}{CE-OFDM}{constant-envelope OFDM}
\newacronym{cept}{CEPT}{European Conference of Postal and Telecommunications Administrations}
\newacronym{cf}{CF}{cell-free}
\newacronym{cfmmimo}{CF-mMIMO}{cell-free massive MIMO}
\newacronym{cfo}{CFO}{carrier frequency offset}
\newacronym{cfra}{CFRA}{contention-free random access}
\newacronym{cir}{CIR}{channel impulse response}
\newacronym{cla}{CLA}{closed-loop approach}
\newacronym{clk}{CLK}{clock}
\newacronym{cmos}{CMOS}{complementary metal oxide semiconductor}
\newacronym{cn}{CN}{Core network}
\newacronym{cnn}{CNN}{convolutional neural network}
\newacronym{co}{CO}{continuous operation}
\newacronym{cordic}{CORDIC}{coordinate rotation digital computer}
\newacronym{cots}{COTS}{commercial off-the-shelf}
\newacronym{cp}{CP}{cyclic prefix}
\newacronym{cpe}{CPE}{common phase error}
\newacronym{cpfsk}{CPFSK}{continuous phase frequency shift keying}
\newacronym{cpm}{CPM}{continuous phase modulation}
\newacronym{cpo}{CPO}{carrier phase offset}
\newacronym{cpt}{CPT}{capacitive power transfer}
\newacronym{cpw}{CPW}{coplanar waveguide}
\newacronym{cqi}{CQI}{channel quality indicator}
\newacronym{cr}{CR}{coding rate}
\newacronym{crc}{CRC}{cyclic redundancy check}
\newacronym{crlb}{CRLB}{Cram\'er-Rao lower bound}
\newacronym{crs}{CRS}{cell reference signal}
\newacronym{crtwt}{C-RTWT}{coordinated restricted TWT}
\newacronym{cs}{CS}{compressed sensing}
\newacronym{cse}{CSE}{channel state estimation}
\newacronym{csi}{CSI}{channel state information}
\newacronym{csp}{CSP}{contact service point}
\newacronym{css}{CSS}{chirp spread spectrum}
\newacronym{cu}{CU}{central unit}
\newacronym{cv}{CV}{constant voltage}
\newacronym{cw}{CW}{continuous wave}
\newacronym{d2d}{D2D}{device-to-device}
\newacronym{d2r}{D2R}{device-to-reader}
\newacronym{dab}{DAB}{Digital Audio Broadcasting}
\newacronym{dac}{DAC}{digital-to-analog converter}
\newacronym{daq}{DAQ}{data acquisition system}
\newacronym{das}{DAS}{distributed antenna systems}
\newacronym{db}{DB}{duty-cycled barebone}
\newacronym{dbpsk}{DBPSK}{Differential Binary Phase Shift Keying}
\newacronym{dc}{DC}{direct current}
\newacronym{dcc}{DCC}{dynamic cooperation clustering}
\newacronym{ddc}{DDC}{digital down conversion}
\newacronym{de}{DE}{drain efficiency}
\newacronym{dect}{DECT}{Digital Cordless Telecommunications}
\newacronym{dft}{DFT}{discrete Fourier transform}
\newacronym{dftsofdm}{DFT-s-OFDM}{discrete Fourier Transform spread OFDM}
\newacronym{dftsofdmfdss}{DFT-s-OFDM-FDSS}{DFT-s-OFDM with frequency-domain spectral shaping}
\newacronym{dftsofdmfdssse}{DFT-s-OFDM-FDSS-SE}{DFT-s-OFDM with FDSS and spectral extension}
\newacronym{dl}{DL}{downlink}
\newacronym{dlc}{DLC}{Distributed Laser Charging}
\newacronym{dli}{DLI}{direct link interference}
\newacronym{dlt}{DLT}{Distributed Ledger Technology}
\newacronym{dm}{DM}{diffuse multipath}
\newacronym{dma}{DMA}{Direct Memory Access}
\newacronym{dmac}{DMAC}{Direct Memory Access Controller}
\newacronym{dmc}{DMC}{diffuse multipath component}
\newacronym{dmz}{DMZ}{demilitarized zone}
\newacronym{dmimo}{D-MIMO}{distributed MIMO}
\newacronym{dnn}{DNN}{deep neural network}
\newacronym{doa}{DOA}{direction-of-arrival}
\newacronym{dp}{DP}{duty-cycled persistent}
\newacronym{dpd}{DPD}{digital pre-distortion}
\newacronym{dpdk}{DPDK}{Data Plane Development Kit}
\newacronym{dram}{DRAM}{dynamic random-access memory}
\newacronym{drcs}{$\Delta$RCS}{differential-radar cross section}
\newacronym{drx}{DRX}{Discontinuous Reception Mode}
\newacronym{dsb}{DSB}{double-sideband}
\newacronym{dsl}{DSL}{digital subscriber line}
\newacronym{dsp}{DSP}{digital signal processing}
\newacronym{dss}{DSS}{Dataset Storage Standard}
\newacronym{duc}{DUC}{digital up-converter}
\newacronym{dvb}{DVB}{Digital Video Broadcasting}
\newacronym{de-ewma}{DE-EWMA}{Dynamic Error-Exponentially Weighted Moving Average}
\newacronym{e2e}{E2E}{end-to-end}
\newacronym{easa}{EASA}{European Union Aviation Safety Agency}
\newacronym{ebg}{EBG}{electromagnetic bandgap}
\newacronym{ebw}{EBW}{excess bandwidth}
\newacronym{ec}{EC}{European Commission}
\newacronym{ecc}{ECC}{Electronic Communications Committee}
\newacronym{ecdf}{eCDF}{empirical cumulative distribution function}
\newacronym{ecdlp}{ECDLP}{elliptic curve discrete logarithm problem}
\newacronym{ecsp}{ECSP}{edge computing service point}
\newacronym{edlc}{EDLC}{electrostatic double-layer capacitors}
\newacronym{edrx}{eDRX}{Extended Discontinuous Reception Mode}
\newacronym{ee}{EE}{energy efficiency}
\newacronym{egprs}{EGPRS}{Enhanced Data Rates for GSM Evolution}
\newacronym{eh}{EH}{energy harvesting}
\newacronym{eirp}{EIRP}{equivalent isotropically radiated power}
\newacronym{em}{EM}{electromagnetic}
\newacronym{embb}{eMBB}{enhanced Mobile Broadband}
\newacronym{en}{EN}{energy neutral}
\newacronym{end}{END}{energy-neutral device}
\newacronym{enob}{ENOB}{effective number of bits}
\newacronym{eol}{EoL}{end of life}
\newacronym{ep}{EP}{energy profiler}
\newacronym{epd}{EPD}{electronic paper display}
\newacronym{epu}{EPU}{edge processing unit}
\newacronym{er}{ER}{Energy Receiver}
\newacronym{erc}{ERC}{European Radiocommunications Committee}
\newacronym{erp}{ERP}{effective radiation power}
\newacronym[plural=ESCs,firstplural=electronic speed controllers (ESCs)]{esc}{ESC}{electronic speed control}
\newacronym{esd}{ESD}{electrostatic discharge}
\newacronym{esl}{ESL}{electronic shelf label}
\newacronym{esr}{ESR}{equivalent series resistance}
\newacronym{et}{ET}{Energy Transmitter}
\newacronym{etsi}{ETSI}{European Telecommunications Standards Institute}
\newacronym{evd}{EVD}{eigenvalue decomposition}
\newacronym{evm}{EVM}{Error Vector Magnitude}
\newacronym{ewlb}{eWLB}{Embedded Wafer Level Ball Grid Array}
\newacronym{ewma}{EWMA}{Exponentially Weighted Moving Average}
\newacronym{eipm}{EIPM}{Energy Intelligence Platform Module}
\newacronym{fa}{FA}{federation anchor}
\newacronym{fair}{FAIR}{Findability, Accessibility, Interoperability, and Reuse of digital assets}
\newacronym{fc}{FC}{fusion center}
\newacronym{fcc}{FCC}{Federal Communications Commission}
\newacronym{fcf}{FCF}{frequency correlation function}
\newacronym{fd}{FD}{front-haul distance}
\newacronym{fdd}{FDD}{frequency-division duplexing}
\newacronym{fde}{FDE}{frequency domain equalizer}
\newacronym{fdm}{FDM}{frequency-division multiplexing}
\newacronym{fdma}{FDMA}{frequency division multiple access}
\newacronym{fdss}{FDSS}{frequency-domain spectral shaping}
\newacronym{fem}{FEM}{finite element analysis}
\newacronym{fembb}{feMBB}{further enhanced mobile broadband}
\newacronym{fft}{FFT}{fast Fourier transform}
\newacronym{fgs}{FGS}{facility geocencing system}
\newacronym{fh}{FH}{fronthaul}
\newacronym{fhss}{FHSS}{frequency hopping spread spectrum}
\newacronym{fifo}{FIFO}{First In, First Out}
\newacronym{fim}{FIM}{Fisher information matrix}
\newacronym{fits}{FITS}{Flexible Image Transport System}
\newacronym{fl}{FL}{Federated learning}
\newacronym{fom}{FoM}{figure of merit}
\newacronym{fov}{FoV}{field of view}
\newacronym{fpga}{FPGA}{field-programmable gate array}
\newacronym{fqt}{FQT}{Fully Quantized Training}
\newacronym{fr1}{FR1}{frequency range 1}
\newacronym{fr2}{FR2}{frequency range 2}
\newacronym{fram}{FRAM}{Ferroelectric Random Access Memory}
\newacronym{fsk}{FSK}{frequency shift keying}
\newacronym{fspl}{FSPL}{free space path loss}
\newacronym{fss}{FSS}{frequency selective surface}
\newacronym{gan}{GaN}{gallium nitride}
\newacronym{gb}{GB}{grant-based}
\newacronym{gdpr}{GDPR}{general data protection regulation}
\newacronym{gf}{GF}{grant-free}
\newacronym{gmsk}{GMSK}{Gaussian minimum-shift keying}
\newacronym{gnb}{gNB}{Next Generation Node B}
\newacronym{gni}{GNI}{gross national income}
\newacronym{gnn}{GNN}{graph neural network}
\newacronym{gnss}{GNSS}{global navigation satellite system}
\newacronym{gpclk}{GPCLK}{general purpose clock}
\newacronym{gpio}{GPIO}{General-Purpose Input/Output}
\newacronym{gpl}{GPL}{GNU General Public License}
\newacronym{gprs}{GPRS}{General Packet Radio Services}
\newacronym{gps}{GPS}{Global Positioning System}
\newacronym{gpu}{GPU}{graphical processing unit}
\newacronym{grc}{GRC}{GNU Radio Companion}
\newacronym{gscm}{GSCM}{geometry‐based stochastic model}
\newacronym{gsm}{GSM}{Global System for Mobile Communications}  %
\newacronym{gspm}{GSpM}{Generalized spatial modulation}
\newacronym{gwp}{GWP}{Global Warming Potential}
\newacronym{har}{HAR}{human activity recognition}
\newacronym{harq}{HARQ}{hybrid automatic repeat request}
\newacronym{hat}{HAT}{hardware attached on top}
\newacronym{hcs}{HCS}{human-centric services}
\newacronym{hdf5}{HDF5}{Hierarchical Data Format version 5}
\newacronym{hdr}{HDR}{high data rate}
\newacronym{hfss}{HFSS}{High Frequency Simulator Software}
\newacronym{hf}{HF}{high frequency}
\newacronym{hmd}{HMD}{head-mounted display}
\newacronym{hpbm}{HPBM}{half power beam width}
\newacronym{HyMPRo}{HyMPRo}{Hybrid Multi-Path Routing algorithm}
\newacronym{i.i.d.}{i.i.d.}{independent and identically distributed}
\newacronym{i2c}{I2C}{Inter-Integrated Circuit}
\newacronym{iaq}{IAQ}{Indoor Air Quality}
\newacronym{ib}{IB}{in-band}
\newacronym{ibbc}{IBBC}{inter-band beam configuration}
\newacronym{ibo}{IBO}{input back-off}
\newacronym{ic}{IC}{integrated circuit}
\newacronym{iccs}{ICCS}{Ilmsens correlative channel sounder}
\newacronym{ici}{ICI}{intercarrier interference}
\newacronym{icnirp}{ICNIRP}{International Commission on Non-Ionizing Radiation Protection}
\newacronym{id}{ID}{information decoding}
\newacronym{idft}{IDFT}{inverse discrete Fourier transform}
\newacronym{idxm}{IDXM}{index modulation}
\newacronym{ieee}{IEEE}{Institute of Electrical and Electronics Engineers}
\newacronym{if}{IF}{intermediate-frequency}
\newacronym{ifft}{IFFT}{inverse fast-Fourier-transform}
\newacronym{iid}{i.i.d.}{independently and identically distributed}
\newacronym{iiot}{IIoT}{Industrial IoT}
\newacronym{iis}{IIS}{integrated information system}
\newacronym{im}{IM}{intermodulation}
\newacronym{imd}{IMD}{intermodulation distortion}
\newacronym{imu}{IMU}{inertial measurement unit}
\newacronym{inh}{InH}{indoor hotspot office}
\newacronym{io}{IO}{input/output}
\newacronym{ioe}{IoE}{Internet of Everything}
\newacronym{iot}{IoT}{Internet of Things}
\newacronym{iotd}{IoTD}{IoT device}
\newacronym{ipt}{IPT}{inductive power transfer}
\newacronym{ipy}{IPY}{Interventions per Year}
\newacronym{iq}{IQ}{in-phase and quadrature}
\newacronym{iqi}{IQI}{IQ imbalance}
\newacronym{ir}{IR}{infrared}
\newacronym{isac}{ISAC}{integrated sensing and communication} 
\newacronym{isi}{ISI}{intersymbol interference}
\newacronym{ism}{ISM}{industrial, scientific and medical}
\newacronym{isp}{ISP}{internet service provider}
\newacronym{itu}{ITU}{International Telecommunication Union}
\newacronym{itu-r}{ITU-R}{ITU – Radiocommunication Sector}
\newacronym{jesd}{JESD}{Joint Electron Devices Engineering Council}
\newacronym{jfet}{JFET}{junction field effect transistor}
\newacronym{kpi}{KPI}{key performance indicator}
\newacronym{ktofdm}{KT-DFT-s-OFDM}{known-tail-DFT-s-OFDM}
\newacronym{kvi}{KVI}{key value indicator}
\newacronym{larva}{LARVA}{LARge Virtual Array}
\newacronym{lbt}{LBT}{Listen-Before-Talk}
\newacronym{lca}{LCA}{life cycle assessment}
\newacronym{lch}{LCH}{logical channel}
\newacronym{lco}{LCO}{lithium cobalt oxide}
\newacronym{ldo}{LDO}{low-dropout voltage regulator}
\newacronym{ldpc}{LDPC}{low-density parity-check}
\newacronym{ldr}{LDR}{low data rate}
\newacronym{led}{LED}{Light Emitting Diode}
\newacronym{less}{LESS}{Low Energy Scheduler Solution}
\newacronym{lev}{LEV}{light electric vehicle}
\newacronym{lfp}{LFP}{lithium iron phosphate}
\newacronym{lib}{LIB}{Lithium-Ion Battery}
\newacronym{lic}{LIC}{lithium-ion capacitor}
\newacronym{lid}{LID}{Lithium Iron Disulfide}
\newacronym{lidar}{LiDAR}{light detection and ranging}
\newacronym{liion}{Li-ion}{lithium-ion}
\newacronym{lipo}{LiPo}{lithium polymer}
\newacronym{lis}{LIS}{large intelligent surface}
\newacronym{llh}{LLH}{log-likelihood}
\newacronym{lls}{LLS}{link-level simulation}
\newacronym{lmd}{LMD}{Lithium Manganese Dioxide}
\newacronym{lmmse}{LMMSE}{least minimum mean square error}
\newacronym{lmo}{LMO}{lithium ion manganese oxide}
\newacronym{lna}{LNA}{low-noise amplifier}
\newacronym{lo}{LO}{local oscillator}
\newacronym{lora}{LoRa}{long range}
\newacronym{lorawan}{LoRaWAN}{long-range wide-area network}
\newacronym{los}{LoS}{line-of-sight}
\newacronym{lp}{LP}{linear programming}
\newacronym{lpf}{LPF}{low-pass filter}
\newacronym{lpt}{LPT}{laser power transfer}
\newacronym{lpwa}{LPWA}{Low Power Wide Area}
\newacronym{lpwan}{LPWAN}{low-power wide-area network}
\newacronym{lpwans}{LPWANs}{Low-Power Wide-Area Networks}
\newacronym{lqi}{LQI}{link quality indicator}
\newacronym{llr}{LLR}{log-likelihood ratio}
\newacronym{lrelu}{LReLU}{leaky rectified linear unit}
\newacronym{lrt}{LRT}{likelihood-ratio test}
\newacronym{ls}{LS}{least squares}
\newacronym{lsa}{LSA}{large synthetic array}
\newacronym{lsf}{LSF}{large-scale fading}
\newacronym{lsfc}{LSFC}{large-scale fading component}
\newacronym{lstm}{LSTM}{Long Short-Term Memory}
\newacronym{ltc}{LTC}{lithium thionyl chloride}
\newacronym{lte}{LTE}{Long Term Evolution}
\newacronym{ltem}{LTE-M}{Long-Term Evolution Machine Type Communication}
\newacronym{lti}{LTI}{linear time-invariant}
\newacronym{lto}{LTO}{lithium titanate}
\newacronym{lusta}{LUSTA 5G }{Logistique mUltimodale Sécuritaire Téléopérée \& Autonome 5G}
\newacronym{m2m}{M2M}{machine to machine}
\newacronym{ma}{MA}{Movable Antennas}
\newacronym{mac}{MAC}{Medium Access Control}
\newacronym{mate}{MATE}{millimeter-wave MIMO testbed}
\newacronym{mc}{MC}{Monte Carlo}
\newacronym{mcl}{MCL}{Maximum Coupling Loss}
\newacronym{mcs}{MCS}{modulation and coding scheme}
\newacronym{mcu}{MCU}{microcontroller unit}
\newacronym{mec}{MEC}{multi-access edge computing}
\newacronym{meh}{MEH}{mobile edge host}
\newacronym{mems}{MEMS}{micro-electromechanical systems}
\newacronym{mf}{MF}{matched filter}
\newacronym{milp}{MILP}{Mixed Integer Linear Programming}
\newacronym{mimo}{MIMO}{multiple-input multiple-output}
\newacronym{miso}{MISO}{multiple-input single-output}
\newacronym{ml}{ML}{machine learning}
\newacronym{mlp}{MLP}{multilayer perceptron}
\newacronym{mmic}{MMIC}{monolithic microwave integrated circuit}
\newacronym{mmimo}{mMIMO}{massive MIMO}
\newacronym{mmse}{MMSE}{minimum mean square error}
\newacronym{mmtc}{mMTC}{massive machine-type communications}
\newacronym{mmwave}{mmWave}{millimeter wave}
\newacronym{mn}{MN}{matching network}
\newacronym{mosfet}{MOSFET}{metal-oxide semiconductor field effect transistor}
\newacronym{mpc}{MPC}{multipath component}
\newacronym{mpp}{MPP}{magnetic power profile}
\newacronym{mppt}{MPPT}{maximum power point tracking}
\newacronym{mqtt}{MQTT}{Message Queuing Telemetry Transport}
\newacronym{mr}{MR}{maximum ratio}
\newacronym{mrc}{MRC}{maximum ratio combining}
\newacronym{mrc_em}{MRC}{maximum ratio combining}
\newacronym{mrc_EM}{MRC}{Magnetic Resonance Coupling}
\newacronym{mrt}{MRT}{maximum ratio transmission}
\newacronym{mse}{MSE}{mean square error}
\newacronym{msk}{MSK}{Minimum-Shift Keying}
\newacronym{mtc}{MTC}{Machine-Type Communication}
\newacronym{multi-rat}{Multi-RAT}{multiple radio access technology}
\newacronym{multirat}{Multi-RAT}{Multiple Radio Access Technology}
\newacronym{music}{MUSIC}{MUltiple SIgnal Classification}
\newacronym{nas}{NAS}{Neural Architecture Search}
\newacronym{navauwall}{NAVAUWALL}{AUtomated NAVigation in WALLonia}
\newacronym{nb}{NB}{narrowband}
\newacronym{nbiot}{NB-IoT}{narrowband IoT}
\newacronym{nca}{NCA}{nickel cobalt aluminum}
\newacronym{netcdf}{NetCDF}{Network Common Data Form}
\newacronym{nf}{NF}{noise figure}
\newacronym{nfc}{NFC}{near-field communication}
\newacronym{nfv}{NFV}{network function virtualization}
\newacronym{ngmn}{NGMN}{Next Generation Mobile Networks }
\newacronym{ni}{NI}{National Instruments}
\newacronym{nicd}{NiCd}{nikkel cadmium}
\newacronym{nimh}{NiMH}{nikkel metal hydride}
\newacronym{nlos}{NLoS}{non-line-of-sight}
\newacronym{nmc}{NMC}{nickel manganese cobalt}
\newacronym{nmos}{nMOS}{n-channel metal-oxide semiconductor}
\newacronym{nn}{NN}{neural network}
\newacronym{nnls}{NNLS}{non-negative least squares}
\newacronym{noma}{NOMA}{non-orthogonal multiple access}
\newacronym{np}{NP}{Neyman-Pearson}
\newacronym{npbch}{NPBCH}{Narrowband Physical Broadcast Channel}
\newacronym{npss}{NPSS}{Narrow Band Primay Synchronization Signal}
\newacronym{nr}{NR}{New Radio}
\newacronym{nrs}{NRS}{Narrow Band Reference Signal}
\newacronym{nsss}{NSSS}{Narrowband Secondary Synchronization Signal}
\newacronym{ntp}{NTP}{network time protocol}
\newacronym{oai}{OAI}{OpenAirInterface} 
\newacronym{obw}{OBW}{occupied bandwidth}
\newacronym{odi}{O-DI}{On-Device Intelligence}
\newacronym{oef}{OEF}{Organisation Environmental Footprint}
\newacronym{ofa}{OFA}{Once-for-All}
\newacronym{ofdm}{OFDM}{orthogonal frequency-division multiplexing}
\newacronym{ofdma}{OFDMA}{orthogonal frequency-division multiple access}
\newacronym{ofdmim}{OFDM-IM}{OFDM with index modulation}
\newacronym{olos}{OLoS}{obstructed-line-of-sight}
\newacronym{oma}{OMA}{orthogonal multiple access}
\newacronym{oob}{OOB}{out-of-band}
\newacronym{ook}{OOK}{on-off keying}
\newacronym{opbo}{OPBO}{output power backoff}
\newacronym{oran}{O-RAN}{open radio-access network}
\newacronym{os}{OS}{operating system}
\newacronym{ota}{OTA}{over-the-air}
\newacronym{otaa}{OTAA}{over-the-air authentication}
\newacronym{otaml}{OTA-TinyML}{Over-the-air TinyML}
\newacronym{ova}{OVA}{One-Versus-All}
\newacronym{p1}{P1}{Phase 1}
\newacronym{p2}{P2}{Phase 2}
\newacronym{p2p}{P2P}{point-to-point}
\newacronym{pa}{PA}{power amplifier}
\newacronym{pae}{PAE}{power-added efficiency}
\newacronym{pam}{PAM}{pulse amplitude modulation}
\newacronym{pana}{PanA}{Panel A}
\newacronym{panb}{PanB}{Panel B}
\newacronym{papr}{PAPR}{peak-to-average power ratio}
\newacronym{pb}{PB}{power beacon}
\newacronym{pc}{PC}{pilot count}
\newacronym{pcb}{PCB}{printed circuit board}
\newacronym{pcg}{PCG}{power consumption gain}
\newacronym{pcie}{PCIe}{Peripheral Component Interconnect Express}
\newacronym{pcsi}{PCSI}{perfect channel state information}
\newacronym{pd}{PD}{powered device}
\newacronym{pdcch}{PDCCH}{physical downlink control channel}
\newacronym{pdcp}{PDCP}{packet data convergence protocol}
\newacronym{pdf}{PDF}{probability density function}
\newacronym{pdp}{PDP}{power delay profile}
\newacronym{pdsch}{PDSCH}{physical downlink shared channel}
\newacronym{pe}{PE}{processing element}
\newacronym{peb}{PEB}{positioning error bound}
\newacronym{pef}{PEF}{Product Environmental Footprint}
\newacronym{per}{PER}{packet error rate}
\newacronym{pet}{PET}{privacy enhancing technology}
\newacronym{peet}{PET}{Polyethylene terephthalate}
\newacronym{pg}{PG}{path gain}
\newacronym{pgd}{PGD}{proximal gradient descent}
\newacronym{phy}{PHY}{physical layer}
\newacronym{pin}{PIN}{positive-intrinsic-negative}
\newacronym{pki}{PKI}{public key infrastructure}
\newacronym{pl}{PL}{path loss}
\newacronym{pla}{PLA}{physically large array}
\newacronym{pla2}{PLA}{polylactic acid}
\newacronym{pll}{PLL}{phase-locked loop}
\newacronym[plural=PMs,firstplural=person months (PMs)]{pm}{PM}{person month}
\newacronym{pmf}{PMF}{polymer microwave fiber}
\newacronym{pmic}{PMIC}{power management integrated circuit}
\newacronym{pmu}{PMU}{power management unit}
\newacronym{pn}{PN}{pseudo-noise}
\newacronym{po}{PO}{phase offset}
\newacronym{poc}{PoC}{proof of concept}
\newacronym{poe}{PoE}{power-over-Ethernet}
\newacronym{pps}{1PPS}{pulse per second}
\newacronym{pr}{PR}{phase reversal}
\newacronym{prb}{PRB}{Physical Resource Block}
\newacronym{prbs}{PRBs}{Physical Resource Blocks}
\newacronym{prs}{PRS}{Peripheral Reflex System}
\newacronym{ps}{PS}{Processing System}
\newacronym{psd}{PSD}{power spectral density}
\newacronym{pse}{PSE}{power sourcing equipment}
\newacronym{psk}{PSK}{phase shift keying}
\newacronym{psm}{PSM}{power saving mode}
\newacronym{pss}{PSS}{primary synchronisation signal}
\newacronym{ptp}{PTP}{precision-time protocol}
\newacronym{ptrs}{PTRS}{Phase-Tracking Reference Signals}
\newacronym{ptw}{PTW}{paging time window}
\newacronym{pv}{PV}{photovoltaic}
\newacronym{pw}{PW}{planar wavefront}
\newacronym{pwm}{PWM}{pulse width modulation}
\newacronym{qam}{QAM}{quadrature amplitude modulation}
\newacronym{qos}{QoS}{quality-of-service}
\newacronym{qpsk}{QPSK}{quadrature phase-shift keying}
\newacronym{qrrls}{QR-RLS}{QR decomposition based recursive least squares}
\newacronym{quadriga}{QuaDRiGa}{QUAsi Deterministic RadIo channel GenerAtor}
\newacronym{r2d}{R2D}{reader-to-device}
\newacronym{ra}{RA}{Random Access}
\newacronym{ram}{RAM}{random-access memory}
\newacronym{ran}{RAN}{radio access network}
\newacronym{rar}{RAR}{Random Access Response}
\newacronym{rat}{RAT}{radio access technology}
\newacronym{raw}{RAW}{restricted access window}
\newacronym{rb}{Rb}{Rubidium}
\newacronym{rbs}{RBS}{radio base station}
\newacronym{rbw}{RBW}{resolution bandwidth}
\newacronym{rc}{RC}{raised-cosine}
\newacronym{rcs}{RCS}{radar cross section}
\newacronym{rdl}{RDL}{redistribution layer}
\newacronym{re}{RE}{radio element}
\newacronym{red}{RED}{radio equipment directive}
\newacronym{redcap}{RedCap}{reduced capability}
\newacronym{relu}{ReLU}{rectified linear unit}
\newacronym{rf}{RF}{radio frequency}
\newacronym{rfeh}{RF-EH}{radio frequency energy harvesting}
\newacronym{rfic}{RFIC}{radio-frequency integrated circuit}
\newacronym{rfid}{RFID}{radio frequency identification}
\newacronym{rfpt}{RFPT}{radio frequency power transfer}
\newacronym{rfsoc}{RFSoC}{Radio Frequency System-on-Chip}
\newacronym{rfwpt}{RF-WPT}{radio frequency wireless power transfer}
\newacronym{rir}{RIR}{room impulse response}
\newacronym{ris}{RIS}{reflective intelligent surface}
\newacronym{rl}{RL}{Reinforcement Learning}
\newacronym{rlc}{RLC}{Radio Link Control}
\newacronym{rllmtc}{RLLMTC}{reliable low latency machine type communication}
\newacronym{rls}{RLS}{recursive least squares}
\newacronym{rms}{RMS}{root-mean-square}
\newacronym{rmse}{RMSE}{root-mean-square error}
\newacronym{rmt}{RMT}{random matrix theory}
\newacronym{rof}{RoF}{radio-over-fiber}
\newacronym{ros}{ROS}{robot operating system}
\newacronym{rpi}{RPi}{Raspberry Pi}
\newacronym{rpl}{RPL}{Routing Protocol for Low power and Lossy Networks}
\newacronym{rrc}{RRC}{Radio Resource Connection}
\newacronym{rreq}{RREQ}{route request packet}
\newacronym{rsrp}{RSRP}{Reference Signals Received Power}
\newacronym{rsrq}{RSRQ}{Reference Signal Received Quality}
\newacronym{rss}{RSS}{received signal strength}
\newacronym{rssi}{RSSI}{received signal strength indicator}
\newacronym{rtc}{RTC}{real time clock}
\newacronym{rtf}{RTF}{reader talks first}
\newacronym{rtk}{RTK}{real time kinematics}
\newacronym{rts}{RTS}{ray tracing simulator}
\newacronym{rtwt}{rTWT}{restricted TWT}
\newacronym{ru}{RU}{Radio Unit}
\newacronym{ruc}{rUC}{representative Use Cases}
\newacronym{rv}{RV}{random variable}
\newacronym{rw}{RW}{RadioWeaves}
\newacronym{rx}{RX}{receiver}
\newacronym{rzf}{RZF}{regularized zero forcing}
\newacronym{s-parameter}{S-parameter}{scattering parameter}
\newacronym{sa}{SA}{system aspects}
\newacronym{sa5}{SA}{Stand Alone}
\newacronym{sar}{SAR}{specific absorption rate}
\newacronym{sbl}{SBL}{sparse Bayesian learning}
\newacronym{sc}{SC}{single carrier}
\newacronym{scomp}{SC}{Split Computing}
\newacronym{scfde}{SC-FDE}{single-carrier modulation with frequency-domain-equalization}
\newacronym{scfdma}{SCFDMA}{single-carrier frequency division multiple access}
\newacronym{scs}{SCS}{sub-carrier spacing}
\newacronym{sdap}{SDAP}{service data adaptation protocol}
\newacronym{sdg}{SDG}{Sustainable Development Goal}
\newacronym{sdm}{SDM}{sigma-delta modulator}
\newacronym{sdma}{SDMA}{spatial-division multiple access}
\newacronym{sdn}{SDN}{software-defined network}
\newacronym{sdof}{SDoF}{sigma-delta over fiber}
\newacronym{sdr}{SDR}{software-defined radio}
\newacronym{se}{SE}{spectral efficiency}
\newacronym{sei}{SEI}{specific emitter identification}
\newacronym{ser}{SER}{symbol-error rate}
\newacronym{sf}{SF}{spreading factor}
\newacronym{sfn}{SFN}{single frequency network}
\newacronym{sfo}{SFO}{sampling frequency offset}
\newacronym{sfp}{SFP}{small form-factor pluggable}
\newacronym{sha}{SHA}{Secure Hash Algorithm}
\newacronym{SigMF}{SigMF}{Signal Metadata Format}
\newacronym{sdo}{SDO}{Standards Development Organization}
\newacronym{simo}{SIMO}{single-input multiple-output}
\newacronym{sinr}{SINR}{signal-to-interference-plus-noise ratio}
\newacronym{siso}{SISO}{single-input single-output}
\newacronym{slam}{SLAM}{simultaneous localization and mapping}
\newacronym{slc}{SLC}{spatial leakage suppression}
\newacronym{slerp}{SLERP}{spherical linear interpolation}
\newacronym{sma}{SMA}{SubMiniature version A}
\newacronym{smc}{SMC}{specular multipath component}
\newacronym{smps}{SMPS}{switched mode power supply}
\newacronym{smt}{SMT}{Surface Mount Technology}
\newacronym{sndr}{SNDR}{signal-to-noise-and-distortion ratio}
\newacronym{snidr}{SNIDR}{signal-to-noise-and-interference-and-distortion ratio}
\newacronym{snir}{SNIR}{signal-to-interference-plus-noise ratio}
\newacronym{snr}{SNR}{signal-to-noise ratio}
\newacronym{soc}{SoC}{state of charge}
\newacronym{SoC}{SOC}{System on Chip}
\newacronym{sota}{SotA}{state of the art}
\newacronym{sp}{SP}{service point}
\newacronym{spdt}{SPDT}{single pole double throw}
\newacronym{spi}{SPI}{Serial Peripheral Interface}
\newacronym{spst}{SPST}{single pole single throw}
\newacronym{sram}{SRAM}{static random-access memory}
\newacronym{srd}{SRD}{short-range device}
\newacronym{srls}{SRLS}{standard recursive least squares}
\newacronym{srs}{SRS}{Sounding Reference Signal}
\newacronym{ssb}{SSB}{synchronisation signal block}
\newacronym{ssd}{SSD}{solid state drive}
\newacronym{ssq}{SSQ}{simulator sickness questionnaire}
\newacronym{sta}{STA}{station}
\newacronym{steam}{STEAM}{science, technology, engineering, the arts, and mathematics}
\newacronym{svd}{SVD}{singular value decomposition}
\newacronym{sw}{SW}{spherical wavefront}
\newacronym{swipt}{SWIPT}{simultaneous wireless information and power transfer}
\newacronym{synce}{SyncE}{Synchronous Ethernet}
\newacronym{ta}{TA}{timing advance}
\newacronym{tau}{TAU}{tracking area update}
\newacronym{tcer}{TCER}{transported to consumed energy ratio}
\newacronym{tcp}{TCP}{Transmission Control Protocol}
\newacronym{tcxo}{TCXO}{temperature compensated crystal oscillator}
\newacronym{tdce}{TD-CE}{time-domain compression and expansion}
\newacronym{tdd}{TDD}{time division duplexing}
\newacronym{tdma}{TDMA}{time division-multiple access}
\newacronym{tdoa}{TDOA}{time-difference-of-arrival}
\newacronym{teg}{TEG}{Thermoelectric Generator}
\newacronym{tsg}{TSG}{Technical Specification Group}
\newacronym{thz}{THz}{Terahertz}
\newacronym{tinyml}{TinyML}{Tiny Machine Learning}
\newacronym{tinytl}{TinyTL}{Tiny Transfer Learning}
\newacronym{tinyol}{TinyOL}{Tiny Online Learning}
\newacronym{tinyrl}{TinyRL}{Tiny Reinforcement Learning}
\newacronym{tl}{TL}{Transfer Learning}
\newacronym{to}{TO}{timing offset}
\newacronym{toa}{TOA}{time-of-arrival}
\newacronym{tof}{ToF}{time-of-flight}
\newacronym{tosm}{TOSM}{through-open-short-match}
\newacronym{tpms}{TPMS}{Tire-Pressure Monitoring System}
\newacronym{trl}{TRL}{technology readyness level}
\newacronym{trp}{TRP}{Transmission Reception Point}
\newacronym{teng}{TENG}{Triboelectric nanogenerators}
\newacronym{tsn}{TSN}{time-sensitive networking}
\newacronym{ttf}{TTF}{tag talks first}
\newacronym{ttff}{TTFF}{Time To First Fix}
\newacronym{tti}{TTI}{transmission time interval}
\newacronym{ttm}{TTM}{time to market}
\newacronym{ttn}{TTN}{The Things Network}
\newacronym{tx}{TX}{transmitter}
\newacronym{twt}{TWT}{Target Wake Time}
\newacronym{uart}{UART}{Universal Asynchronous Receiver/Transmitter}
\newacronym{uav}{UAV}{unmanned aerial vehicle}
\newacronym{uc}{UC}{use case}
\newacronym{ucie}{UCIe}{Universal Chiplet Interconnect Express}
\newacronym{udp}{UDP}{User Datagram Protocol}
\newacronym{ue}{UE}{user equipment}
\newacronym{ugv}{UGV}{unmanned ground vehicle}
\newacronym{uhd}{UHD}{USRP hardware driver}
\newacronym{uhf}{UHF}{ultra-high frequency}
\newacronym{ul}{UL}{uplink}
\newacronym{ula}{ULA}{uniform linear array}
\newacronym{ule}{ULE}{ultra low energy}
\newacronym{ulp}{ULP}{ultra-low power}
\newacronym{uMIMO}{$\mu$-MIMO}{ultra-massive Multiple-Input Multiple-Output}
\newacronym{ummtc}{umMTC}{ultra massive machine type communication}
\newacronym{un}{UN}{United Nations}
\newacronym{unow}{uNOW}{unified non-orthogonal waveform}
\newacronym{upa}{UPA}{uniform planar array}
\newacronym{up}{UP}{ultra passive}
\newacronym{ura}{URA}{uniform rectangular array}
\newacronym{urllc}{URLLC}{ultra-reliable low-latency communications}
\newacronym{usrp}{USRP}{universal software radio peripheral}
\newacronym{uv}{UV}{unmanned vehicle}
\newacronym{uw}{UW}{unique word}
\newacronym{uwb}{UWB}{ultrawideband}
\newacronym{uwofdm}{UW-DFT-s-OFDM}{unique word DFT-s-OFDM}
\newacronym{v2v}{V2V}{vehicle-to-vehicle}
\newacronym{vco}{VCO}{voltage-controlled oscillator}
\newacronym{vep}{VEP}{virtual edge platform}
\newacronym{vlc}{VLC}{visible light communication}
\newacronym{vlp}{VLP}{visible light positioning}
\newacronym{vna}{VNA}{vector network analyzer}
\newacronym{voc}{VOC}{Voltatile Organic Compound}
\newacronym{votable}{VOTable}{Virtual Observatory Table}
\newacronym{vr}{VR}{virtual reality}
\newacronym{vuca}{VUCA}{volatile, uncertain, complex and ambiguous}
\newacronym{vswr}{VSWR}{Voltage Standing Wave Ratio}
\newacronym{wifi}{Wi-Fi}{Wireless Fidelity}
\newacronym{wirelesshart}{WirelessHART}{Wireless Highway Addressable Remote Transducer Protocol}
\newacronym{wur}{WuR}{wake-up radio}
\newacronym{wus}{WuS}{wake-up signal}
\newacronym{wb}{WB}{wideband}
\newacronym{wban}{WBAN}{wireless body area network}
\newacronym{wd}{WD}{Wireless distance}
\newacronym{wimax}{WiMAX}{Worldwide Interoperability for Microwave Access}
\newacronym{wlan}{WLAN}{wireless LAN}
\newacronym[plural=WPs,firstplural=work packages (WPs)]{wp}{WP}{work package}
\newacronym{wpc}{WPC}{Wireless Power Consortium}
\newacronym{wpt}{WPT}{wireless power transfer}
\newacronym{wr}{WR}{White Rabbit}
\newacronym{wrsn}{WRSN}{wireless rechargeable sensor network}
\newacronym{wsn}{WSN}{Wireless Sensor Network}
\newacronym{wcma}{WCMA}{weather conditioned moving average}
\newacronym{xets}{XETS}{cross exponentially tapered slot}
\newacronym{xlmimo}{XL-MIMO}{extremely large-scale MIMO}
\newacronym{xr}{XR}{extended reality}
\newacronym{z3ro}{Z3RO}{zero third-order distortion}
\newacronym{zed}{ZED}{Zero-Energy device}
\newacronym{zf}{ZF}{zero-forcing}
\newacronym{zmcscg}{ZMCSCG}{zero mean circularly symmetric complex Gaussian}
\newacronym{zmq}{ZMQ}{ZeroMQ}
\newacronym{ztofdm}{ZT-DFT-s-OFDM}{zero-tail DFT-s-OFDM}

\newacronym{mobc}{MoBC}{monostatic backscatter communication}

\newacronym{bibc}{BiBC}{bistatic backscatter communication}
\newacronym{t2t}{T2T}{tag-to-tag}
\newacronym{pls}{PLS}{physical layer security}

\newacronym{cea}{CEA}{controlled environment agriculture}
\newacronym{abd}{ABD}{ambient-backscattering device}
\newacronym{keh}{KEH}{kinetic energy harvester}
\newacronym{svm}{SVM}{support vector machine}
\newacronym{wisp}{WISP}{Wireless Identification and Sensing Platform}

\begin{document}
%
\title{Frugal Geofencing via Energy-aware  \\Sensing and Reporting}
\author{\IEEEauthorblockN{1\textsuperscript{st} David E. Ru\'{i}z-Guirola}
\IEEEauthorblockA{\textit{CWC, University of Oulu}\\
Oulu, Finland \\
David.RuizGuirola@oulu.fi}
\and
\IEEEauthorblockN{2\textsuperscript{nd} 
Miltiadis C. Filippou}
\IEEEauthorblockA{\textit{WINGS ICT Solutions S.A.} \\
Athens, Greece \\
mfilippou@wings-ict-solutions.eu}
\and
\IEEEauthorblockN{3\textsuperscript{rd} Onel L. A. L\'{o}pez}
\IEEEauthorblockA{\textit{CWC, University of Oulu}\\
Oulu, Finland \\
Onel.AlcarazLopez@oulu.fi}

\thanks{This work has been partially supported by the Research Council of Finland (Grants 369116 (6G Flagship) and 362782 (ECO-LITE)), the Finnish Foundation for Technology Promotion, and the European Commission through the Horizon Europe/JU SNS project AMBIENT-6G (Grant 101192113).}
}

%
%

\markboth{Journal of \LaTeX\ Class Files,~Vol.~14, No.~8, August~2015}%
{Shell \MakeLowercase{\textit{et al.}}: Bare Demo of IEEEtran.cls for IEEE Journals}
%



\maketitle

\begin{abstract}
Timely and accurate monitoring in geofencing scenarios is challenging when relying on ultra-low power Internet of Things devices (IoTDs) powered by energy harvesting (EH). This is mainly because frequent wake-ups for data acquisition and data uploading may quickly deplete their limited energy buffer. Conventional grid-like IoT deployments overlook these limitations and merely rely on continuously powered sensing.
Herein, we propose an energy-aware geofencing framework for camera-equipped EH IoTDs deployed around a protected area and its surrounding perimeter zone. The framework integrates a directional sensing power model with an operational representation of EH, sensing, sleeping, and reporting, accounting for the limited field-of-view (FoV) and distance-dependent detection confidence of the IoTDs. Device activity is controlled by the coverage-providing access point, which hosts a mobile edge host and a facility geocencing system to ensure timely and reliable detection under tight energy constraints. Reinforcement learning is used to determine IoTD placement, enabling earlier intruder detection than uniform grid-based deployments. Numerical results show that the proposed coordinated sensing and reporting configuration achieves frugal geofencing with fewer devices, while concurrently improving detection timeliness and dependability.
\end{abstract}


%
\IEEEpeerreviewmaketitle

\section{Introduction}
\IEEEPARstart{T}{he} \gls{iot} is a key element of modern digital transformation, allowing for the seamless integration of physical objects into the digital ecosystem. It connects a variety of \glspl{iotd} equipped with sensors, actuators, computing, and communication modules, enabling them to collect, exchange, and respond to data across different environments~\cite{ruiz2025context}. However, with projections exceeding 38 billion connected \glspl{iotd} by 2029, the prevailing reliance on non-rechargeable batteries becomes increasingly incompatible with scalability and sustainability, driving frequent maintenance interventions and higher operational costs~\cite{ericsson2023}. In this sense, sustainable development increasingly depends on technological solutions that balance efficient resource management and reliable access to essential and critical services~\cite{lopez2023energy}. 

In this context, \gls{aiot} envisions large-scale deployments of battery-free or battery-assisted \glspl{iotd} operating at $\mu$W-level average power. Herein \gls{eh} is combined with a rechargeable energy buffer to enable sporadic sensing/processing/reporting of events with decades-long autonomy~\cite{3gppTS38391, famaey2026survey,lopez2024zero}. However, one single low-power node is not able to guarantee continuous monitoring. Instead, network-wide collaborative sensing is required, where many intermittently active \glspl{iotd} jointly ensure adequate spatial and temporal coverage of the region of interest~\cite{he2022collaborative}. As deployments expand and underpin safety and mission-critical infrastructure, stringent requirements on availability, reliability, and latency further intensify. This makes the design of energy-aware sensing and reporting policies a central problem for dependable \gls{aiot} operation~\cite{ruiz2025context}. 


Barrier coverage is a key aspect of collaborative sensing for access control along a boundary, where a moving target must be detected by the barrier formed by the \glspl{iotd}. Recent works have introduced quality-aware methods to capture performance degradation in harsh environments and sensor faults~\cite{he2022collaborative,he2021efficient}. Deployment strategies, including grid-based and application-specific designs, can reduce the sensor density required to achieve high reliability with respect to random deployment~\cite{yang2024trajectory}. When sensing is directional, such as with cameras, full-view barrier coverage further constrains the problem by requiring each point of the barrier to be observed under specific viewing conditions, motivating deployment strategies to minimize the number of cameras while ensuring complete coverage~\cite{he2022collaborative,shi2022toward,du2024full}. 
To address coverage gaps caused by offsets, overlapping views, or failures, prior works have considered sensor mobility~\cite{li2022optimization}, optimal \gls{iotd} selection~\cite{shi2022toward}, and camera rotatability or mobility~\cite{du2024full}. In parallel, energy-aware scheduling techniques have been studied to extend network lifetime while maintaining coverage quality, including balancing coverage and energy consumption, selective \gls{iotd} activation, and sleep-wake scheduling~\cite{chowdary2021pre,chang2022energy,sivakumar2023enhancing}. 

Despite these advances, most existing approaches assume that \glspl{iotd} can operate continuously or with relatively high duty cycles. In general, they do not account for the intermittency of the \glspl{iotd} operating with \gls{eh}, directional sensor \gls{fov} constraints, and/or the joint design of sensing and reporting schedules within a unified \gls{aiot} framework. Thus, ensuring reliable barrier coverage with energy-constrained \glspl{iotd} remains a pending challenge~\cite{ruiz2026IoTJ}.  

To maintain performance under strict energy budgets, \glspl{iotd} and/or the network should intelligently decide when to sleep, sense, and report. Thus, an intelligent deployment must be complemented with dynamic duty cycling and adaptive sensing schedules to sustain network availability. Herein, we focus on an \gls{iotd}-based geofencing deployment for accurate and early intrusion detection with frugal use of on-device energy, signaling, and computing resources. 
The main contributions of this paper are as follows:
\begin{itemize}[leftmargin=8pt,align=left]
  \item We present a geofencing system for low-power camera-equipped \glspl{iotd}, where a protected area (or asset) is surrounded by an outer perimeter zone that allows for early intruder detection before they cross the boundary. 
  \item We propose a directional sensing‑power function that fuses distance decay and sensor (camera) \gls{fov} masking to quantify object‑detection confidence over time. We also model energy-aware operation, including sensing, sleeping, reporting, and stochastic \gls{eh} dynamics. 
  \item We formulate a geofenced area protection problem in which the \gls{ap} provides geofencing system control by dynamically adjusting device activity, reporting, and sensing orientation to balance detection reliability, timeliness, and energy sustainability.
  \item We assess the proposed intelligent and exploration-based \gls{iotd} deployments (applying a \gls{rl}-based approach) and sensing configurations. We demonstrate how target mobility-aware control facilitated by \gls{ap} control signaling can decrease the necessary device density (and, therefore, system cost), while maintaining high reliability and timely detection. 
\end{itemize}
Section~\ref{sec:2_system} describes the system model, network architecture, geofencing and perimeter zone setup, signaling aspects, energy model, and directional sensing. Section~\ref{sec:3_problem} formulates the protected area access control problem and presents the proposed optimization approach. Section~\ref{sec:results} provides the numerical results along with a discussion of the performance. Finally, we conclude the paper in Section~\ref{sec:conclusions}.

\section{System Model}\label{sec:2_system}

We consider the geofencing scenario shown in \figurename~\ref{fig:scenarios}, where a radio \gls{ap} provides coverage to a set of $N$ \glspl{iotd} deployed in an area of interest, such as a warehouse or industrial site. Spatially, the monitored setup consists of two concentric regions: (i) a protected geofenced area, which represents the facility or perimeter that needs to be secured, and (ii) an outer perimeter zone that acts as an early detection buffer around it. The deployed \glspl{iotd} are arranged so that their directional sensing sectors collectively create a barrier around the geofenced area, while also extending some coverage into the perimeter zone. The outer perimeter zone serves as an early-warning area where approaching intruders are detected first. This enhances the chances of timely identification and classification before they enter the protected area at time $t_{\mathrm{g}}^{(o)}$. This capability allows for proactive alarms starting from the moment an intruder enters the outer perimeter at time $t_{\mathrm{in}}^{(o)}$ and enables earlier adjustments to device duty cycles and reporting decisions.
\begin{figure}
    \centering
    \includegraphics[width=0.95\linewidth]{fig/System_model4.png}
    \caption{System model for energy-aware geofencing intrusion detection. A set of camera-equipped \glspl{iotd} with directional sensing capability is deployed around the protected geofenced area, while an outer zone provides an early detection region for approaching intruders before boundary crossing. The \gls{ap}-\gls{meh} collects local detection reports (red dashed line), predicts trajectories, and dynamically adapts sensing and reporting decisions (wake-up requests, green dashed line) according to task relevance and device energy availability.}
    \label{fig:scenarios}
\end{figure}

Each \gls{iotd} is equipped with an optical camera
of a given \gls{fov} ($\Delta\theta$, assumed identical for all \gls{iotd} cameras), vision depth and rotation capability, a \gls{eh} module, an energy buffer, and a \gls{wur}. Moreover, it includes application-related hardware, such as a data storage element which, apart from a data buffer, hosts a lightweight Computer Vision (CV) model which is trained to detect different object classes (e.g., humans, vehicles, warehouse inventory, etc.), along with a \gls{rf} communication module for the needed communication with the \gls{ap} and a low-power microcontroller. 


During \gls{iotd} operation, the CV model is used in inference mode, ingesting the locally generated camera video frames. A MEH of given processing, memory, and storage resource capability is deployed in the vicinity of the \gls{ap}. The MEH hosts a \gls{fgs}, composed of a Controller that regulates the operation of the \glspl{iotd} towards fulfilling the geofencing goal in an energy-aware manner. To effectively control the deployment's operation, we assume that the Controller receives information by each deployed \gls{iotd}, either periodically (i.e., the energy buffer level), or on an event-based fashion (e.g., video feed interruption and restoration time). Apart from the Controller, the \gls{fgs} incorporates a \gls{fc} to obtain and process the local detection output data from all \glspl{iotd}, when they are uploaded via the \gls{ap}. In terms of network deployment flexibility, the \gls{ap} is assumed to be permanently fixed at a particular location, whereas the \glspl{iotd} can be easily placed at different locations of the area of interest to best address the geofence entry detection requirements. 

Detected objects are reported to the \gls{ap} together with some metadata, including \gls{iotd} status information, useful to the Controller. Based on such uplink signaling to the \gls{ap}, the received information reaches the \gls{meh}, where the \gls{fgs} is hosted. Then, within the \gls{fgs}, the \gls{fc} fuses detection reports and predicts remaining intruder trajectories, and, subsequently triggers the Controller to reconfigure the most relevant \glspl{iotd} according to task relevance and energy availability. As a result, the Controller instructs the \gls{ap} to issue event-driven wake-up requests to selected \glspl{iotd} through their \gls{wur} interface. These requests are triggered when an approaching intruder is detected, and the predicted trajectory indicates that additional sensing support is needed from sleeping \glspl{iotd}. In this way, only the most relevant \glspl{iotd} are activated, which improves timely detection, while avoiding unnecessary energy consumption. 

\subsection{Sensing power function}

\begin{figure}[t!]
    \centering
    \includegraphics[width=0.8\linewidth]{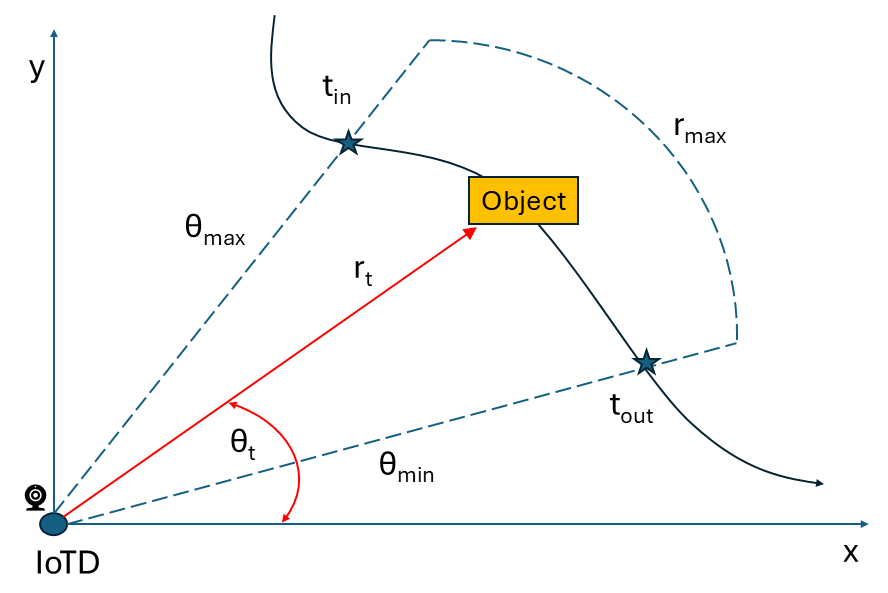}
    \caption{Detection sector of an \gls{iotd} (e.g., camera) showing the object trajectory, entry and exit points, along with the angular and radial constraints.}
    \label{fig:detection-sector}
\end{figure}

\figurename~\ref{fig:detection-sector} shows an \gls{iotd} with an (azimuth) angular sector of $\Delta\theta = \theta_{\max} - \theta_{\min}$ degrees, which is determined by both $\theta_{\min}$ and the \gls{fov}. Additionally, the figure includes an object with an already performed (up to time instant $t$) and planned trajectory that needs to be detected, as well as the involved geometrical and time parameters that will be explained in what follows.
Assuming that time is measured in \glspl{tti}, the sensing function $p(r_t, \theta_t)$ captures the detection confidence of the target within one observation window (\gls{tti}), under the assumption that the target remains fixed during that interval.

Since the targeted object is not detectable when present at an azimuth angle outside the camera \gls{fov} of the given \gls{iotd}, i.e., outside the angular interval $[\theta_{min}, \theta_{max}]$, we need to consider the following angular masking function
\begin{equation}
    I(\Delta\theta_t) = 
    \begin{cases}
        1, & \text{if $\theta_o \in [\theta_{min}, \theta_{max}]$},\\
        0, & \text{otherwise}.
    \end{cases}
\end{equation}
Given that the targeted object is within the \gls{iotd} camera \gls{fov}, we propose a decaying function of the distance between the \gls{iotd} and the targeted object, which nulls at distance $r_{max}$. This captures the detection capability of the \gls{iotd}, which may be based, for example, on a CV algorithm applied for object recognition. Such algorithms result in a maximum number of false negatives (practically not detecting any object) when these targeted objects are located at distances beyond a maximum distance within which detection is feasible. Herein, for a given \gls{tti} period, we assume that the distance-dependent sensing function is modeled as follows
\begin{equation}
    p(r_t) = 
    \begin{cases}
        \exp{(-\eta r_t}), & \text{if $r_t \in [0, r_{\text{max}}]$},\\
        0, & \text{otherwise}.
    \end{cases}
\end{equation}
Thus, taking into account both distance and angular domains (equivalently, working with polar coordinates), the sensing power function for a given time interval of duration of one \gls{tti} is given by
\begin{equation}
    p(r_t, \Delta\theta_t) = p(r_t) I(\Delta\theta_t).
\end{equation}
The detection probability for an object trajectory entering the detectability area of a given \gls{iotd} at time instance $t_{in}$ and leaving it at time instance $t_{out}$, when measured at time $t \in [t_{in}, t_{out}]$, is
\begin{equation}
p_{\mathrm{obj}}=\max_{t\in[t_{\mathrm{in}},t_{\mathrm{out}}]} p(r_{t},\Delta\theta_{t}).
\end{equation}
It should be noted that, for a given \gls{iotd}, in case of non-detection of an object for the time when the object is within the detectability area of the \gls{iotd}, the detection task concludes unsuccessfully, i.e., the context of this task is not retained until the previously non-detected object re-enters the \gls{iotd} camera \gls{fov}. In the latter (object return) case, the \gls{iotd} will launch a new detection task.

\subsection{Successful object detection}

Upon detecting an intruder, the \glspl{iotd} upload a detection update to the \gls{ap}, optionally accompanied by a classification label (e.g., ``human'', ``vehicle'', or ``animal''). The \gls{fc} collects the updates from the \glspl{iotd} and declares a successful object detection as soon as at least one valid detection report is received. A detection report is deemed valid when the sensing confidence satisfies the detection condition $p_{\mathrm{obj}} \geq p_{\mathrm{th}}$. After that point, the network no longer needs to adapt sensing, wake-up, and uplink reporting decisions to that object's motion, since the object has already been detected and identified.
If continuous object tracking until departure from the protected area is also required, then the sensing and reporting policy should continue adapting to the target trajectory over time. However, extending the problem from detection and recognition to continuous tracking is outside the scope of this paper and is left for future work.

\subsection{Energy considerations} 

We adopt a normalized, discrete \gls{eh} model with battery capacity $C_{\max}$ and independent Bernoulli energy arrivals for each \gls{iotd}. At each \gls{tti}, each \gls{iotd} independently harvests up to $P_{\mathrm{B}}$ units with probability $\lambda$, and harvested energy is stored up to $C_{\max}$, while overflow is discarded.\footnote{The proposed model can adapt to different \gls{eh} sources (e.g., solar, \gls{rf}). In that case, probability $\lambda$ can be modeled as dependent on context parameters, for instance, the time-of-day, the \gls{iotd} distance from the \gls{rf} signaling source, etc. Such modeling and its accuracy verification are left for future work.} On the energy consumption side, transmission, wake-up operation, and camera reorientation are actions consuming $P_{\mathrm{Tx}}$, $P_{\mathrm{WuR}}$, and $P_{\mathrm{rot,bin}}$ units per \gls{tti}, respectively. If the energy buffer level is insufficient for the selected action, the action is unsuccessful, and the device remains unavailable until enough energy is accumulated again.

\section{Problem formulation \& proposed solution}\label{sec:3_problem} 

The goal of the \gls{fgs} is to detect each target as early as possible within the outer perimeter zone, ideally before it crosses the geofenced boundary. Let $t_{\mathrm{in}}^{(o)}$ denote the time at which intruder $o$ first enters the outer zone, $t_{\mathrm{g}}^{(o)}$ the time at which it crosses the geofenced boundary, and $t_{\mathrm{det}}^{(o)}$ the time of the first valid detection report for that intruder. For each intruder $o$, detection is classified as early if {${t_{\mathrm{det}}^{(o)}<t_{\mathrm{g}}^{(o)}}$}, late if {$t_{\mathrm{det}}^{(o)}\ge t_{\mathrm{g}}^{(o)}$}, and missed if no valid detection report is produced while the intruder remains within the monitored region.

We define the object-level detection error as
\begin{equation}
E_{\mathrm{obj}}(o)= 
    \begin{cases}
        0, & \text{early detection},\\
        \mu_1, & \text{late detection},\\
        1, & \text{misdetection},
    \end{cases}
\label{eq:Eobj}
\end{equation}
where $\mu_1\in[0,1]$ penalizes late detection less severely than a complete miss. Note that $E_{\mathrm{obj}}(o)$ is evaluated during training and simulation using ground-truth object trajectories. In deployment, the AP executes the learned policy online based only on the available network state information. 

Maximizing timely detection while minimizing undetected intrusions depends on the energy availability, position, and \gls{fov} orientation of the deployed \glspl{iotd}. Activating more \glspl{iotd} and increasing their duty cycles can reduce blind spots, but it also increases sensing, reporting, and reconfiguration energy consumption. Conversely, aggressive energy saving may leave parts of the perimeter insufficiently monitored or prevent low-energy \glspl{iotd} from reporting valid detections. Therefore, the design objective is to jointly configure sensing activity, event-driven wake-up, and camera orientation to minimize missed and late detections while preserving long-term device availability. 

\subsection{Problem formulation}\label{sec:opt}

Let $\boldsymbol{\delta}(t)=\left[\delta_1(t),\delta_2(t),\ldots,\delta_N(t)\right]$ denote the \gls{fgs}-issued sensing policy to be applied at \gls{tti} $t$, where $\delta_j(t)\in\{0,1\}$ indicates whether \gls{iotd} $j$ should be (remain or change its state to, if different) inactive/sleeping or active/sensing, respectively. Given this policy, the \gls{ap} issues event-driven wake-up requests to selected \glspl{iotd}, the camera \glspl{fov} of which are predicted to, at least, partly cover the remaining intruder trajectory. Likewise, $\boldsymbol{\Delta\theta}(t)=\left[\Delta\theta_1(t),\Delta\theta_2(t),\ldots,\Delta\theta_N(t)\right]$ denote the \gls{fov} orientation vector of the deployed \glspl{iotd}. Since camera reorientation is energy-consuming, $\boldsymbol{\Delta\theta}(t)$ is updated less frequently than $\boldsymbol{\delta}(t)$. 

Then, the joint \gls{iotd} activation and camera rotation optimization problem can be stated as
\begin{equation}
    \min_{\boldsymbol{\delta}(t),\,\boldsymbol{\Delta\theta}(t)}
    \frac{1}{\alpha T}\sum_{\forall o} E_{\mathrm{obj}}(o),
\label{eq:P1_new}
\end{equation}
where $T$ is the total number of analyzed \glspl{tti} and $\alpha$ is the expected number of intruders per \gls{tti}. The goal is to optimize the activation and orientation policies to minimize the average intruder detection error, favoring early detection and penalizing late detections and complete misses. 

\subsection{\gls{rl}-based proposal}\label{sec:rl}

The \gls{fgs} must repeatedly adapt its decisions to time-varying object trajectories, partial sensing outcomes, and \gls{eh} conditions. For this reason, we adopt a centralized \gls{rl}-based approach, where the \gls{fgs} Controller acts as the decision agent and dynamically selects the sensing states and camera orientations of the most relevant \glspl{iotd} in order to balance system availability and detection reliability and timeliness. In its capacity, the \gls{ap} acts as a policy enforcement entity.

At \gls{tti} $t$, the system state is defined as $\mathcal{T}(t)=\left\{\mathcal{T}_j(t)\right\}_{j\in\mathcal{N}}$, with per-device state
\begin{equation}
\mathcal{T}_j(t)=
\left\{
\delta_j(t),\,
\Delta\theta_j(t),\,
p(r_t,\Delta\theta_t),\,
(x_j,y_j),\,
B_j(t)
\right\},
\label{eq:states_RL_new}
\end{equation}
where $(x_j,y_j)$ is the position of \gls{iotd} $j$, and $B_j(t)$ is its energy buffer level at time $t$. 
Given the observed state $\mathcal{T}(t)$, the \gls{fgs} Controller selects the activation vector $\boldsymbol{\delta}(t)$ and, when required, updates the orientation vector $\boldsymbol{\Delta\theta}(t)$. Hence, the learned policy is ${\pi^*:\mathcal{T}(t)\rightarrow \left(\boldsymbol{\delta}(t),\boldsymbol{\Delta\theta}(t)\right),}$
to maximize the long-term expected reward. 
Moreover, to transform the desired objective in~\eqref{eq:P1_new} to the \gls{rl} domain, we define the reward function as
\begin{equation}
r(t)=
\left(1-\frac{\boldsymbol{\delta}(t)\mathbf{1}^{\mathsf T}}{N}\right)
+\mu_2\mathrm{Cov}(t)
-\alpha\mu_3 E_{\mathrm{obj}}(t),
\label{eq:RL_reward_new}
\end{equation}
where $\mathrm{Cov}(t)$ denotes the fraction of the perimeter effectively covered by active \glspl{iotd}, and $\mu_2,\mu_3>0$ are weighting factors. The first term promotes energy efficiency by penalizing unnecessary activations, the second rewards spatial coverage, and the third penalizes late detections and missed detections through the intruder error in~\eqref{eq:Eobj}. In this way, the \gls{fgs}, via the \gls{ap} learns to activate and orient only the most trajectory-relevant \glspl{iotd}, while preserving energy resources for sustained geofencing operation.

\section{Results analysis}\label{sec:results}
In the geofencing scenario considered in this paper, we focus on two major \glspl{kpi}: (i) reliability, meaning that systems with active \glspl{iotd} must detect intruders with probability at least $99.9\%$, and (ii) timeliness, per which detections should occur sufficiently early to enable countermeasures before the target enters the protected area. These requirements directly expose the fundamental trade-off discussed in Section~\ref{sec:3_problem}, where increasing sensing activity improves detection timeliness, but also increases energy consumption and may lead to prolonged device unavailability due to depleted energy buffers~\cite{3gppTS38391,ruiz2025context}. 

\subsection{Simulation setup}


In this work, we evaluate two deployment/control strategies:
\textbf{(i)} a grid baseline with static placement and periodic sleep/active operation and 
\textbf{(ii)} the proposed \gls{rl}-based mobility-aware configuration, where the \gls{fgs} jointly controls device activation $\boldsymbol{\delta}(t)$ and camera orientation $\boldsymbol{\theta}(t)$. 
The sensing periodicity is parameterized by the duty-cycle period $\tau$ (in ms), spanning $\tau\in\{1, 4, 8,\cdots, 1024\}$ ms. These values represent feasible operating points derived from \gls{aiot} use cases, considering different mean and peak power constraints as well as corresponding energy-neutral sensing frequency limits~\cite{3gppTS38391,famaey2026survey}. For each pair of parameters $(N, \tau)$, we simulate multiple independent intruder trajectories and compute the following metrics: the successful detection rate $(P_{\mathrm{det}})$, the early detection probability $(P_{\mathrm{early}})$ as the likelihood that detection occurs with sufficient lead time before boundary crossing, and the mean time to detect $(\mathbb{E}[T_{\mathrm{det}}])$. 



We assume a normalized discrete energy and sensing model. Each \gls{iotd} has battery capacity $C_{\max}=10$ units and harvests energy with $P_{\mathrm{B}}=1$ unit and probability $\lambda=0.1$~\cite{ruiz2026IoTJ,ruiz2024intelligent}. Moreover, $P_{\mathrm{Tx}}=1$ unit, $P_{\mathrm{WuR}}=0.01$ units, and $P_{\mathrm{rot,bin}}=0.1$ units per $30^{\circ}$ camera reorientation~\cite{jinlong2023wisecam}. The sensing setup assumes \gls{fov} $=60^{\circ}$~\cite{sirmacek2020occupancy}, $r_{\max}=3$ m, $\eta=1$, and sensing confidence $p_{\mathrm{th}}=\exp(-0.5)$~\cite{ruiz2026IoTJ,ruiz2024intelligent}. The protected geofence area is $7\times7$ m$^2$, surrounded by an outer zone of $8\times8$ m$^2$, while $N\in[4,10^5]$, and the intruder speed is fixed to 1 $m/s$. Additionally, we model intruder arrivals by a time-varying Poisson process with higher intensity during working hours at 10 arrivals per hour from 06:00 to 18:00, and a lower intensity at night with 0.5 arrivals per hour.

\subsection{Detection reliability versus \gls{iotd} density and duty cycle}
\begin{figure}[t!]
\centering
\includegraphics[width=\linewidth]{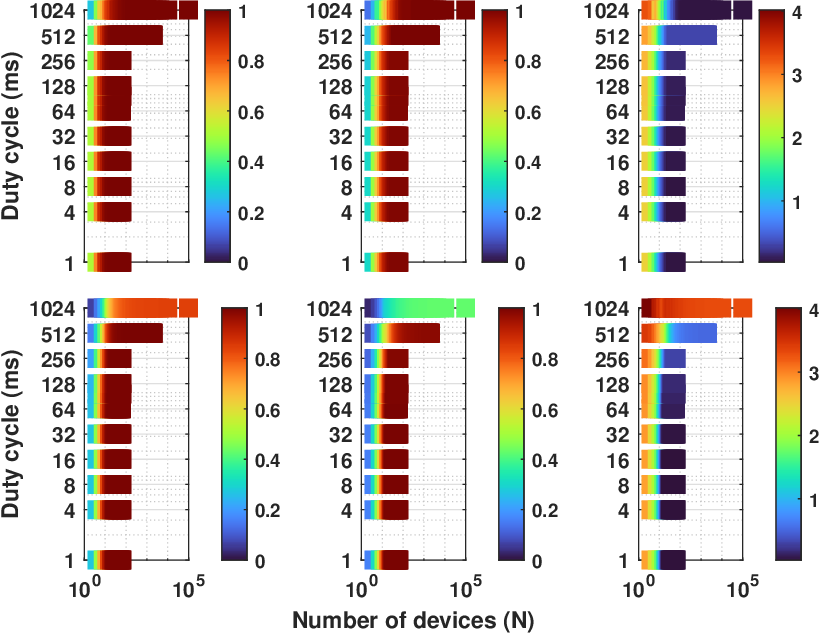}
\caption{Impact of \gls{iotd} density $N$ and duty-cycle period $\tau$ on reliability and timeliness for the geofencing intrusion-detection task. The bottom row corresponds to the grid baseline and the top row to the proposed \gls{rl}-based mobility-aware configuration. Columns report (left) detection rate $P_{\mathrm{det}}$, (center) early-detection probability $P_{\mathrm{early}}$, and (right) mean time-to-detect $\mathbb{E}[T_{\mathrm{det}}]$ (s).}
\label{fig:heatmaps_all}
\end{figure}

\figurename~\ref{fig:heatmaps_all} reports the three metrics as a function of \gls{iotd} density $N$ (x-axis) and duty-cycle period $\tau$ (y-axis). Therein, we compare the \gls{rl}-based approach (top) and the grid baseline (bottom). As expected, for any fixed $\tau$, both approaches exhibit a reliability that increases with $N$,  rapidly driving $P_{\mathrm{det}}$ over $99.9\%$, as the intruder is more likely to intersect at least one \gls{iotd}’s \gls{fov} during an active interval. However, the \gls{rl}-based policy systematically shifts this transition to the left, needing smaller $N$ for moderate-to-large duty cycles, indicating that mobility-aware activation/orientation decisions improve the effective sensing exposure of the network per unit energy. 

\begin{figure}[t!]
\centering
\includegraphics[width=0.33\linewidth]{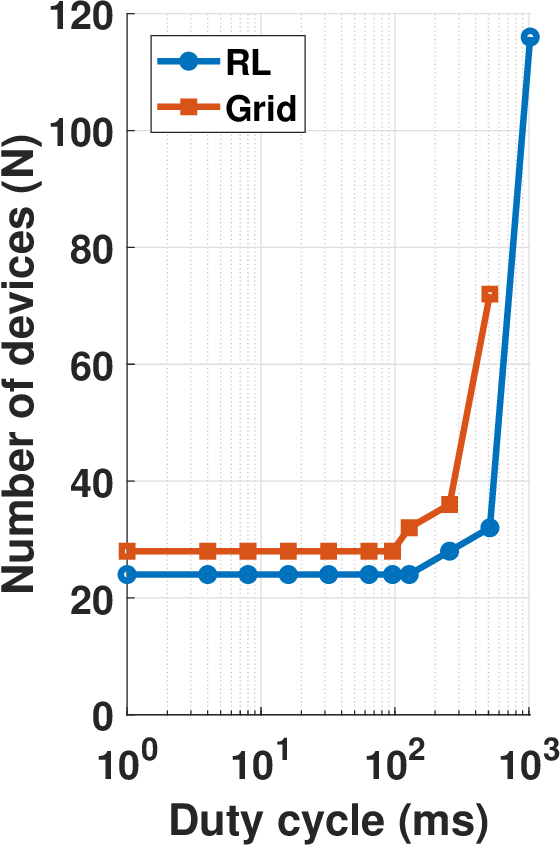}
\includegraphics[width=0.65\linewidth]{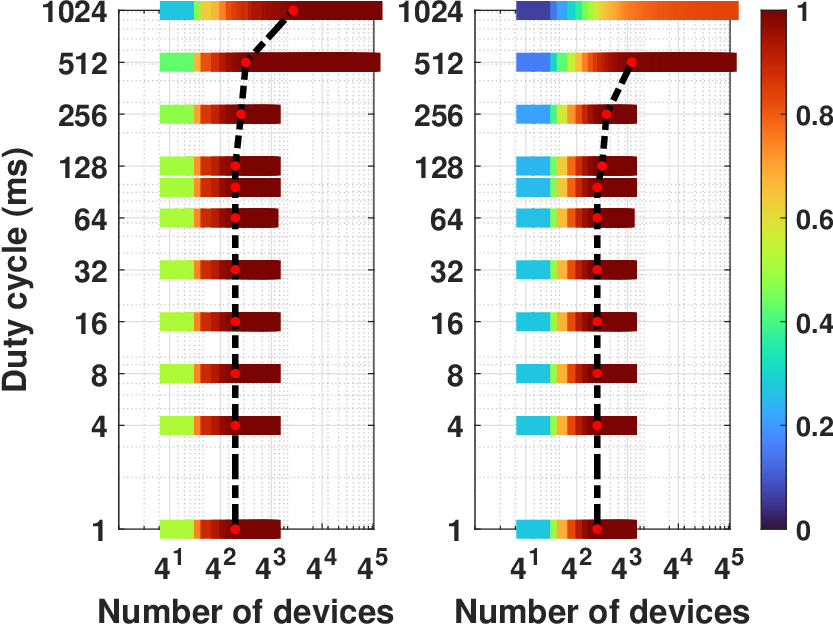}
\caption{Minimum required \gls{iotd} density $N_{\min}(\tau)$ to satisfy $P_{\mathrm{det}}\ge 0.999$ reliability operating point as a function of the duty-cycle period (left). Detection-rate map with the $99.9\%$ reliability frontier required for active \gls{aiot} operation for the proposed \gls{rl}-based proposal (center) and the grid baseline (right). The black dashed curve with red markers indicates $N_{\min}(\tau)$, the minimum \gls{iotd} density needed to achieve $P_{\mathrm{det}}(N,\tau)\ge 0.999$ for each duty-cycle period $\tau$, same as the left plot.}
\label{fig:det_boundary}
\end{figure}

\figurename~\ref{fig:det_boundary} overlays the $99.9\%$ reliability boundary (black dashed curve with red markers indicating the minimum number of \glspl{iotd} to reach $99.9\%$ of reliability), defined as
\begin{align}
    N_{\min}(\tau) \triangleq \min{\{N;P_{\mathrm{det}}(N,\tau)\ge 0.999\}}
\end{align}
which is aligned with the active \gls{aiot} reliability target in~\cite{3gppTS38391}. Notably, for tight duty cycles (small $\tau$), both methods reach $0.999$ reliability with relatively low $N_{\min}$ because frequent sensing reduces the chance of ``missing'' the intruder in time. In contrast, as $\tau$ grows (energy-constrained regimes), the grid baseline requires a rapidly increasing number of \glspl{iotd} to maintain $P_{\mathrm{det}}\ge 0.999$, whereas the \gls{rl}-based approach needs fewer \glspl{iotd}, reflecting its ability to re-allocate sensing effort towards the most trajectory-relevant \glspl{iotd} and their orientations under the same energy constraints. This is summarized in \figurename~\ref{fig:det_boundary}, where the \gls{rl}-based approach achieves the target reliability with substantially fewer \glspl{iotd} than the grid one in the medium/high-$\tau$ region, thus reducing deployment cost while preserving the \gls{kpi}. 

\subsection{Early detection and latency at the 99.9\% reliability}

Reliability alone is insufficient for intrusion prevention. Late detections, even if eventually successful, may not allow enough time to trigger protective actions. Therefore, we next evaluate the timeliness at the reliability-feasible operating point $N=N_{\min}(\tau)$. \figurename~\ref{fig:heatmaps_all} shows $P_{\mathrm{early}}$ (center) and $\mathbb{E}[T_{\mathrm{det}}]$ (right) under the grid method and the \gls{rl}-based approach at $N_{\min}$. For small $\tau$, the grid method shows slightly lower $P_{\mathrm{early}}$ and mean detection delay, which is consistent with frequent periodic sensing combined with uniform perimeter coverage. Nevertheless, as $\tau$ increases, the grid method timeliness degrades markedly. Longer sleep intervals create larger temporal sensing gaps, increasing both the probability of detecting ``too late'' and the expected detection delay. 
In contrast, the \gls{rl}-based approach sustains timeliness more robustly as $\tau$ increases. This behavior is explained by the coupling between mobility-awareness and energy-awareness. By prioritizing \glspl{iotd} that are closer to the predicted intruder route and adjusting \gls{fov} to reduce blind spots, the \gls{rl}-based approach increases the probability that an intruder is observed soon after arrival despite longer sleep cycles. Moreover, the reliability-constrained choice $N_{\min}(\tau)$ implicitly captures a second-order trade-off: at very large $\tau$, meeting $P_{\mathrm{det}}\ge0.999$ requires larger $N$, and this added spatial redundancy can partially compensate for sparse sensing, yielding improved $P_{\mathrm{early}}$ and reduced $\mathbb{E}[T_{\mathrm{det}}]$ even under stringent energy budgets.

\subsection{\gls{iotd} Availability}

Table~\ref{tab:energy_summary} presents the average energy buffer levels of the deployed \glspl{iotd} at various times throughout the day for both the \gls{rl} and grid methods. Note that the \gls{rl}-based policy maintains significantly higher energy levels across nearly all \glspl{iotd} throughout the day. In contrast, the grid method often leads many \glspl{iotd} to critically low or depleted energy states. Therefore, this means that the \gls{rl} approach more efficiently distributes sensing and reporting actions, preventing excessive energy drain on individual \glspl{iotd} and ensuring a larger number of \glspl{iotd} remain operational over time. Consequently, the \gls{rl} method enhances device availability, which directly contributes to more reliable barrier coverage and improves the continuity of protection against intrusions.
\begin{table}[t]
\caption{Energy levels over 20 \glspl{iotd}. Devices with energy below 15\% are considered unavailable for reliable protection.}
\label{tab:energy_summary}
\centering
\setlength{\tabcolsep}{3pt}
\renewcommand{\arraystretch}{1.03}
\begin{tabular}{c|cc|cc|cc}
\hline
Time & \multicolumn{2}{c|}{Average energy (\%)} & \multicolumn{2}{c|}{Available \glspl{iotd}} & \multicolumn{2}{c}{Depleted \glspl{iotd}} \\
 & RL & Grid & RL & Grid & RL & Grid \\
\hline
06:00 & 37.13 & 16.38 & 20 & 15 & 0 & 4 \\
09:00 & 29.09 & 11.54 & 19 & 10 & 1 & 5 \\
12:00 & 25.37 & 7.56  & 19 & 5  & 1 & 6 \\
15:00 & 22.00 & 5.88  & 16 & 1  & 1 & 10 \\
18:00 & 20.12 & 4.40  & 16 & 0  & 1 & 9 \\
21:00 & 52.94 & 30.48 & 20 & 13 & 0 & 0 \\
\hline
\end{tabular}
\end{table}

\section{Conclusion}\label{sec:conclusions}
This paper introduced a geofencing framework for camera-equipped \gls{eh} \glspl{iotd}. By combining a directional sensing model, an energy-aware operation model, and \gls{rl}-based control, the approach facilitated earlier and more reliable intrusion detection within the outer perimeter while using fewer \glspl{iotd} than a static grid. Results showed that the RL method lowers the necessary \gls{iotd} density, improves detection timeliness, and ensures higher energy availability across \glspl{iotd}.


%




\ifCLASSOPTIONcaptionsoff
  \newpage
\fi

\bibliographystyle{IEEEtran}
\bibliography{short_bib}

\begin{thebibliography}{10}
\providecommand{\url}[1]{#1}
\csname url@samestyle\endcsname
\providecommand{\newblock}{\relax}
\providecommand{\bibinfo}[2]{#2}
\providecommand{\BIBentrySTDinterwordspacing}{\spaceskip=0pt\relax}
\providecommand{\BIBentryALTinterwordstretchfactor}{4}
\providecommand{\BIBentryALTinterwordspacing}{\spaceskip=\fontdimen2\font plus
\BIBentryALTinterwordstretchfactor\fontdimen3\font minus \fontdimen4\font\relax}
\providecommand{\BIBforeignlanguage}[2]{{%
\expandafter\ifx\csname l@#1\endcsname\relax
\typeout{** WARNING: IEEEtran.bst: No hyphenation pattern has been}%
\typeout{** loaded for the language `#1'. Using the pattern for}%
\typeout{** the default language instead.}%
\else
\language=\csname l@#1\endcsname
\fi
#2}}
\providecommand{\BIBdecl}{\relax}
\BIBdecl

\bibitem{ruiz2025context}
D.~E. Ruiz-Guirola \emph{et~al.}, ``{Context-awareness for Dependable Low-Power IoT},'' \emph{arXiv preprint arXiv:2510.23125}, 2025.

\bibitem{ericsson2023}
P.~Jonsson, ``{Ericsson Mobility Report},'' Ericsson, Tech. Rep., 2023.

\bibitem{lopez2023energy}
O.~L.~A. L{\'o}pez \emph{et~al.}, ``{Energy-Sustainable IoT Connectivity: Vision, Technological Enablers, Challenges, and Future Directions},'' \emph{IEEE Open J. Commun. Soc.}, vol.~4, pp. 2609--2666, 2023.

\bibitem{3gppTS38391}
``{3GPP TS 38.391: NR; Ambient IoT Medium Access Control (MAC) Protocol},'' ETSI (3GPP), Tech. Rep. TS 138 391, 2025.

\bibitem{famaey2026survey}
J.~Famaey \emph{et~al.}, ``{A Survey on Ambient IoT: Use Cases, Standardization, Enabling Technologies, and Research Directions},'' \emph{Authorea Preprints}, 2026.

\bibitem{lopez2024zero}
O.~L{\'o}pez \emph{et~al.}, ``{Zero-Energy Devices for 6G: Technical Enablers at a Glance},'' \emph{IEEE Internet Things Mag.}, vol.~8, no.~3, pp. 14--22, 2025.

\bibitem{he2022collaborative}
S.~He \emph{et~al.}, ``{Collaborative Sensing in Internet of Things: A Comprehensive Survey},'' \emph{IEEE Commun. Surveys Tuts.}, vol.~24, no.~3, pp. 1435--1474, 2022.

\bibitem{he2021efficient}
------, ``{Efficient Fault-Tolerant Information Barrier Coverage in Internet of Things},'' \emph{IEEE Trans. Wireless Commun.}, vol.~20, no.~12, pp. 7963--7976, 2021.

\bibitem{yang2024trajectory}
F.~Yang and L.~Shu, ``{A Trajectory-Inspired Node Deployment Strategy in Solar Insecticidal Lamps Internet of Things Under Coverage and Maintenance Cost Considerations},'' \emph{IEEE Trans. AgriFood Electron.}, vol.~2, no.~1, pp. 28--42, 2024.

\bibitem{shi2022toward}
K.~Shi \emph{et~al.}, ``{Toward Optimal Deployment for Full-View Point Coverage in Camera Sensor Networks},'' \emph{IEEE Internet Things J.}, vol.~9, no.~21, pp. 22\,008--22\,021, 2022.

\bibitem{du2024full}
H.~Du \emph{et~al.}, ``{Full View Maximum Coverage of Camera Sensors: Moving Object Monitoring},'' \emph{ACM Trans. Sens. Netw.}, vol.~20, no.~3, pp. 1--23, 2024.

\bibitem{li2022optimization}
H.~Li \emph{et~al.}, ``{Optimization Method of Linear Barrier Coverage Deployment for Multistatic Radar},'' \emph{J. Syst. Eng. Electron.}, vol.~34, no.~1, pp. 68--80, 2022.

\bibitem{chowdary2021pre}
V.~Chowdary \emph{et~al.}, ``{Pre-Deployment Strategy for Maximizing Barrier Coverage in Wireless Sensor Network},'' \emph{Int. J. Sens. Wireless Commun. Control}, vol.~11, no.~2, pp. 181--188, 2021.

\bibitem{chang2022energy}
J.~Chang \emph{et~al.}, ``{Energy-Efficient Barrier Coverage Based on Nodes Alliance for Intrusion Detection in Underwater Sensor Networks},'' \emph{IEEE Sensors J.}, vol.~22, no.~4, pp. 3766--3776, 2022.

\bibitem{sivakumar2023enhancing}
N.~R. Sivakumar \emph{et~al.}, ``{Enhancing Network Lifespan in Wireless Sensor Networks Using Deep Learning Based Graph Neural Network},'' \emph{Phys. Commun.}, vol.~59, p. 102076, 2023.

\bibitem{ruiz2026IoTJ}
D.~E. Ru{\'i}z-Guirola \emph{et~al.}, ``{Energy Management and Wake-up for IoT Networks Powered by Energy Harvesting},'' \emph{IEEE Internet Things J.}, pp. 1--1, 2026.

\bibitem{ruiz2024intelligent}
D.~E. Ruiz-Guirola \emph{et~al.}, ``{Intelligent Duty Cycling Management and Wake-up for Energy Harvesting IoT Networks with Correlated Activity},'' in \emph{58th Asilomar Conf. Signals, Syst. Comput.}, 2024.

\bibitem{jinlong2023wisecam}
E.~Jinlong \emph{et~al.}, ``{WiseCam: Wisely Tuning Wireless Pan-Tilt Cameras for Cost-Effective Moving Object Tracking},'' in \emph{IEEE INFOCOM}.\hskip 1em plus 0.5em minus 0.4em\relax IEEE, 2023, pp. 1--10.

\bibitem{sirmacek2020occupancy}
B.~Sirmacek and M.~Riveiro, ``{Occupancy Prediction Using Low-Cost and Low-Resolution Heat Sensors for Smart Offices},'' \emph{Sensors}, vol.~20, no.~19, p. 5497, 2020.

\end{thebibliography}










\end{document}